\documentclass[10pt,prd,aps,twocolumn,preprintnumbers, showpacs, nofootinbib,superscriptaddress,notitlepage]{revtex4-2}
\usepackage{amssymb,amsthm,amsmath}
\usepackage{textcomp}
\usepackage{subfig,graphicx}   
\usepackage{color}      
\usepackage{slashed}    
\usepackage{verbatim}
\usepackage[normalem]{ulem}
\usepackage{rotating}   
\usepackage{multirow}  
\usepackage[colorlinks,
linkcolor=red,
anchorcolor=blue,
citecolor=blue
]{hyperref}

\begin{document}

	\baselineskip=15pt
	
	\preprint{CTPU-PTC-26-18}

\title{
Unraveling weak radiative hyperon decays with broken flavor symmetry
}

\author{Kaiwen Chen}
\affiliation{Department of Physics and Institute of Theoretical Physics,
Nanjing Normal University, Nanjing, Jiangsu 210023, China}
\affiliation{Nanjing Key Laboratory of Particle Physics and Astrophysics,
Nanjing, Jiangsu 210023, China}

\author{Lin Qiu}
\affiliation{Department of Physics and Institute of Theoretical Physics,
Nanjing Normal University, Nanjing, Jiangsu 210023, China}
\affiliation{Nanjing Key Laboratory of Particle Physics and Astrophysics,
Nanjing, Jiangsu 210023, China}

\author{Jin Sun}
\email{sunjin0810@ibs.re.kr}
\affiliation{Particle Theory and Cosmology Group, Center for Theoretical Physics of the Universe, Institute for Basic Science (IBS), Daejeon 34126, Korea }

\author{Zhi-Peng Xing}
\email{zpxing@nnu.edu.cn}
\affiliation{Department of Physics and Institute of Theoretical Physics,
Nanjing Normal University, Nanjing, Jiangsu 210023, China}
\affiliation{Nanjing Key Laboratory of Particle Physics and Astrophysics,
Nanjing, Jiangsu 210023, China}

\author{Ruilin Zhu}
\email{rlzhu@njnu.edu.cn}
\affiliation{Department of Physics and Institute of Theoretical Physics,
Nanjing Normal University, Nanjing, Jiangsu 210023, China}
\affiliation{Nanjing Key Laboratory of Particle Physics and Astrophysics,
Nanjing, Jiangsu 210023, China}

\begin{abstract}

We present a broken $SU(3)$ flavor analysis of  weak radiative decays of spin-1/2 hyperons, incorporating current-current and electromagnetic-penguin contributions as well as charge- and mass-insertion effects. Six independent reduced amplitudes describe all decay channels, while a minimal relation between the parity-conserving and parity-violating form factors reduces  to eight free parameters.
A global fit to the ten available observables yields $\chi^2/\mathrm{d.o.f.}=0.98$ and accommodates the large negative asymmetry in $\Sigma^+\to p\gamma$. The resulting nonzero effective form factor $g_{b_8}$ provides a possible flavor-symmetry realization of the parity-violating amplitude while remaining compatible with Hara’s theorem, which constrains the current-current contribution in the exact symmetry limit. 
Our analysis favors a sizable negative value of
$\alpha_\gamma(\Xi^- \to \Sigma^- \gamma)$, which differs from the current
experimental result by approximately $1.3\sigma$. Future precision
measurements of this observable will provide a decisive test of this
prediction.

\end{abstract}

\maketitle

\noindent{\bf \textit{Introduction} }
Recently, the BESIII Collaboration reported absolute measurements of the branching fractions of several non-leptonic $\Sigma^+$ decays \cite{BESIII:2025rgd} as
\begin{eqnarray}
&&{\cal B}(\Sigma^+\to p\pi^0)=(49.79 \pm 0.06 \pm 0.22)\%,\notag\\
&&{\cal B}(\Sigma^+\to n\pi^+)=(49.87 \pm 0.05 \pm 0.29)\%.
 \end{eqnarray}
These measurements differ significantly from the Particle Data Group (PDG) averages~\cite{PDG},
by 4.4 $\sigma$ for $\Sigma^+\to p\pi^0$ and 3.4 $\sigma$ for $\Sigma^+\to n\pi^+$.
The former tension is consistent with our previous flavor-symmetry analysis \cite{Wu:2025hnh}, in which the PDG value for $\Sigma^+\to p\pi^0$ was disfavored by the remaining decay channels. Updating the global fit with the BESIII results together with the PDG inputs removes this discrepancy and yields
 $\chi^2/d.o.f=0.462$. 
 This agreement further supports the effectiveness of systematically broken $SU(3)$ flavor symmetry in describing hyperon decays and motivates its application to the more challenging case of hyperon radiative decays.

Hyperon radiative decays probe the interplay among electromagnetic, weak, and strong interactions through weak strangeness-changing transitions accompanied by photon emission.
Despite their simple two-body kinematics, these  decays have long posed significant challenges to both theory and experiment\cite{Gavela:1980bp,Nardulli:1987ub,Jenkins:1992ab,Neufeld:1992np,Lach:1995we,Borasoy:1999nt,Dubovik:2008zz,Li:2016tlt,Bigi:2017eni,Xing:2023jnr,Erben:2025jkq}.

Experimentally, the absolute branching fractions and asymmetry parameters $\alpha_\gamma$ of hyperon radiative decays have been measured at several facilities~\cite{PDG}.
BESIII has recently improved the precision for multiple channels~\cite{BESIII:2022rgl,BESIII:2023fhs,BESIII:2024lio,BESIII:2025roh}. Most notably, its 2023 measurement of ${\cal B}(\Sigma^+\to p\gamma)$ deviates from the PDG value  by $4.2\sigma$~\cite{BESIII:2023fhs}. 
These experimental developments call for a renewed theoretical analysis of hyperon radiative decays.

Theoretically, for spin-1/2 hyperons, the decay is governed by a parity-conserving (P-wave) amplitude and a parity-violating (S-wave) amplitude, which determine the matrix element $\mathcal{M}$, width $\Gamma$, and asymmetry parameter $\alpha_\gamma$ as~\cite{PDG}
\begin{align}
\mathcal{M}
&=\frac{G_F e}{M_i+M_f}\,
\bar u_f\, i\sigma_{\mu\nu}q^\nu\epsilon^\mu
\left(F+G\gamma_5\right)u_i,\label{ff}
\\
\Gamma&=
 G_F^2e^2\frac{|\bm q|^3}{\pi(M_i+M_f)^2}
 (|F|^2+|G|^2),\notag\\
\alpha_\gamma&=
 \frac{2\operatorname{Re}(F^*G)}{|F|^2+|G|^2},
 \quad |\bm q|=\frac{M_i^2-M_f^2}{2M_i},\notag
\end{align}
where $M_i$ ($M_f$) is the mass of the initial (final) baryon, while $F$ and $G$ denote the parity-conserving and parity-violating amplitudes, respectively. Within this  framework, the long-standing puzzle is exemplified by the unexpectedly large measured  asymmetry, $\alpha_\gamma(\Sigma^+\to p\gamma)=-0.69\pm0.05$~\cite{PDG}, which contradicts Hara’s theorem~\cite{Hara:1964zz}. 
In the exact $SU(3)$ flavor symmetry limit, the theorem requires the parity-violating amplitudes of the charged decay modes to vanish. In particular, $G_{\rm cc}^{(0)}(\Sigma^+\to p\gamma)=0$,
  where the superscript $(0)$ denotes the exact symmetry limit and the subscript $cc$ denotes the current-current contribution. Consequently, the photon asymmetry parameter also vanishes, $\alpha_\gamma(\Sigma^+\to p\gamma)=0$. 
Beyond the tension with Hara’s theorem, other frameworks, including chiral perturbation theory ($\chi$PT), also struggle to accommodate the observed pattern~\cite{Shi:2025xkp}. 
A fully consistent theoretical description of the available data  remains lacking~\cite{Bos:1996ig,Dubovik:2008zz,Niu:2020aoz,Shi:2022dhw,Shi:2023kbu,Xing:2023jnr}.

Within a flavor symmetry framework, a nonzero parity-violating amplitude may be generated by both $SU(3)$ breaking effects and electromagnetic-penguin contributions~\cite{Zenczykowski:2005cs,Wang:2020wxn}. This motivates a systematic $SU(3)$ analysis incorporating both sources.

In this Letter, we perform a broken $SU(3)$ analysis of  hyperon radiative decays.
By combining the effects of the current-current and electromagnetic-penguin interactions with symmetry-breaking terms, we show that six independent reduced amplitudes suffice to describe all decay channels. 
Under an economical factorization ansatz, the resulting eight-parameter fit successfully describes the ten available observables, with $\chi^2/d.o.f.=0.98$, while reproducing the large negative asymmetry. Our framework therefore provides a minimal and unified flavor description of hyperon radiative decays and yields correlated predictions for future measurements.

\noindent{\bf \textit{$SU(3)$ amplitudes 
}}
Weak radiative hyperon decays can be induced by the penguin-level $s\to d\gamma$ and tree-level $s\to d u\bar u$ Hamiltonians. At the quark level, the relevant low-energy $\Delta S=1$ Hamiltonian may be written as~\cite{Buchalla:1995vs}
\begin{eqnarray}
 {\cal H}_{\rm eff}=\frac{G_F}{\sqrt2}V_{us}V_{ud}^*
 \sum_{i=1-7}\bigl[z_i(\mu)+\tau y_i(\mu)\bigr]Q_i(\mu)
 +{\rm h.c.},
 \label{eq:interaction}
\end{eqnarray}
where $\tau=-V_{ts}V_{td}^*/(V_{us}V_{ud}^*)$. The dominant
current-current and electromagnetic-penguin operators are expressed as
\begin{align} \label{eq:Lagrangian}
 Q_1=&[\bar u_\alpha\gamma_\mu(1-\gamma_5)s_\beta]
 [\bar d_\beta\gamma^\mu(1-\gamma_5)u_\alpha],\notag\\
 Q_2=&[\bar u_\alpha\gamma_\mu(1-\gamma_5)s_\alpha]
 [\bar d_\beta\gamma^\mu(1-\gamma_5)u_\beta],\nonumber\\
  Q_{7\gamma}=&\frac{e}{16\pi^2}m_s\,
 \bar d\,\sigma^{\mu\nu}(1+\gamma_5)s\,F_{\mu\nu},
\end{align}
where QCD-penguin operators $Q_{3\text{--}6}$ are ignored.
The current-current operators $Q_{1,2}$ produce
the tree $W$-exchange flavor structure, whereas $O_{7\gamma}$ produces the penguin tensor.
Following the irreducible-representation-amplitude (IRA) method in Ref.~\cite{Wu:2025hnh}, these Hamiltonians can be represented by an octet and a 27-${\rm plet}$ as
\begin{eqnarray}
Q_{1,2}: (H_8)^2_{3}&=&\frac14\sin\theta_C,\;\;(H_{27})^{12}_{13}=(H_{27})^{12}_{31}=\frac15\sin\theta_C,
 \nonumber\\
 (H_{27})^{22}_{23}&=&(H_{27})^{23}_{33}=-\frac1{10}\sin\theta_C,  \notag
 \end{eqnarray}
 \begin{eqnarray}
Q_{7\gamma}:    (\widetilde H_8)^2_{3}&=&\sin\theta_C=4(H_8)^2_{3},
\end{eqnarray}
where $\theta_C$ denotes the Cabibbo angle.
Note that the symmetric 27-plet Hamiltonian 
is suppressed due to the antisymmetric color structure of the baryon wave function~\cite{Korner:1970xq,Pati:1970fg}. Therefore we omit its contribution in our work.
Within the framework of $SU(3)$ flavor symmetry, the initial- and final-state baryons are represented in terms of an octet as 
\begin{equation}
 T_8=\begin{pmatrix}
 \dfrac{\Sigma^0}{\sqrt2}+\dfrac{\Lambda}{\sqrt6}&\Sigma^+&p\\
 \Sigma^-&-\dfrac{\Sigma^0}{\sqrt2}+\dfrac{\Lambda}{\sqrt6}&n\\
 \Xi^-&\Xi^0&-\dfrac{2\Lambda}{\sqrt6}
 \end{pmatrix}.
\end{equation}

The $SU(3)$-invariant decay amplitudes are then constructed by contracting the remaining flavor indices as
\begin{align}
{\cal{M}}_S=&a(T_8)_i^{j}
(H_8)_j^{k}(T_8)_k^{i} +b(T_8)_i^{k}(H_8)_j^{i}
(T_8)_k^{j}\notag\\
+&\tilde{a}(T_8)_i^{j}
(\widetilde H_8)_j^{k}(T_8)_k^{i} +\tilde{b}(T_8)_i^{k}(\widetilde H_8)_j^{i}(T_8)_k^{j},
 \label{eq:amplitude}
\end{align}
where $a$ and $b$ correspond to $Q_{1,2}$, while $\tilde{a}$ and $\tilde{b}$ correspond to $Q_{7\gamma}$. 
Although evaluating the weak radiative  amplitudes induced by
$Q_{1,2}$ requires   additional photon-emission
operators, their contributions vanish in the exact $SU(3)_F$
symmetry limit, in which the constituent quarks are degenerate~\cite{Wang:2020wxn}. 
It is straightforward to verify that the amplitudes associated with
$\tilde{a}$ and $\tilde{b}$ are identical to those associated with $a$ and $b$. The decay amplitudes can therefore be expressed in terms of only two independent amplitudes as
\begin{align}
{\cal{M}}_S=&a(T_8)_i^{j}
(H_8)_j^{k}(T_8)_k^{i} +b(T_8)_i^{k}(H_8)_j^{i}
(T_8)_k^{j}.
\end{align}

However, once taking into account the quark charges and masses, the degeneracy is lifted and explicit $SU(3)$ breaking is introduced.
These effects  can be represented by 
 a photon-emission operator and a quark-mass insertion, each transforming as   $3\otimes\bar 3=8\oplus 1$, where the singlet component is trivial. The charge-octet matrix $Q$ and mass-difference-octet matrix $M$ are 
\begin{align}
Q=\operatorname{diag}\left(\frac23,-\frac13, -\frac13\right),\; M=\operatorname{diag}\left(-\frac13,-\frac13,\frac23\right).
\end{align}
Using these matrices, the symmetry-breaking amplitude can be constructed as
\begin{align}
{\cal M}_B&=a^1_{m}(T_8)_i^{l}(H_8)_j^{k}(T_8)_k^{i} M_l^{j} +b^1_{m}(T_8)_i^{k}(H_8)_j^{l}(T_8)_k^{j}M_l^{i}\notag\\
&+a^2_{m}(T_8)_i^{j}(H_8)_j^{l}(T_8)_k^{i} M_l^{k} +b^2_{m}(T_8)_i^{k}(H_8)_j^{i}(T_8)_k^{l}M_l^{j}\notag\\
&+a^3_{m}(T_8)_i^{j}(H_8)_j^{k}(T_8)_k^{ l}M_l^{i}+b^3_{m}(T_8)_i^{l}(H_8)_j^{i}(T_8)_k^{j}M_l^{k}\notag\\
&+a^1_{q}(T_8)_i^{l}(H_8)_j^{k}(T_8)_k^{i} Q_l^{j} +b^1_{q}(T_8)_i^{k}(H_8)_j^{l}(T_8)_k^{j}Q_l^{i}\notag\\
&+a^2_{q}(T_8)_i^{j}(H_8)_j^{l}(T_8)_k^{i} Q_l^{k} +b^2_{q}(T_8)_i^{k}(H_8)_j^{i}(T_8)_k^{l}Q_l^{j}\notag\\
&+a^3_{q}(T_8)_i^{j}(H_8)_j^{k}(T_8)_k^{ l}Q_l^{i}+b^3_{q}(T_8)_i^{l}(H_8)_j^{i}(T_8)_k^{j}Q_l^{k},
\end{align}
where the mixed insertion  $Q^i_{\;l}M^l_{\;j}$, encoding  the combined electric charge and quark mass effects, is omitted because it introduces no  independent flavor structure: $Q M = -\frac{1}{9} I - \frac{1}{3} Q - \frac{1}{3} M$.
Its contribution can therefore be absorbed into those associated with
$I$, $Q$, and $M$. The amplitudes $a^1(b^1)$ and $a^2(b^2)$ 
describe the symmetry-breaking insertions on the quark lines connecting  the initial and
final baryons to the effective Hamiltonian, whereas $a^3(b^3)$
denotes the corresponding spectator quark contribution. Since the Hamiltonian flavor structure  is fixed by the underlying
$s \to d$ transition, the terms 
$a^{1,2}$ ($b^{1,2}$) take same flavor structures as $a$ ($b$) and can be redefined as
\begin{align}
a_8 &=
a-\frac{1}{3}\left(a_q^1+a_q^2\right)
+\frac{2}{3}a_m^1-\frac{1}{3}a_m^2, 
\notag\\
b_8 &=
a-\frac{1}{3}\left(b_q^1+b_q^2\right)
+\frac{2}{3}b_m^1-\frac{1}{3}b_m^2,\notag\\
a_{8q}&=a^3_q,\;a_{8m}=a^3_m,\;b_{8q}=b^3_q,\;b_{8m}=b^3_m.
\end{align}
Therefore, 
the total decay amplitude can be parametrized in terms of six independent
amplitudes as
\begin{align}\label{eq:total}
\mathcal{M}
&=\mathcal{M}_S+\mathcal{M}_B\\
&=a_8 (T_8)_i^{\,j}(H_8)_j^{\,k}(T_8)_k^{\,i}
+b_8 (T_8)_i^{\,k}(H_8)_j^{\,i}(T_8)_k^{\,j}
\notag\\
&
+a_{8m} (T_8)_i^{\,j}(H_8)_j^{\,k}
       (T_8)_k^{\,l}M_l^{\,i}
+b_{8m} (T_8)_i^{\,l}(H_8)_j^{\,i}
       (T_8)_k^{\,j}M_l^{\,k}
\notag\\
&
+a_{8q} (T_8)_i^{\,j}(H_8)_j^{\,k}
       (T_8)_k^{\,l}Q_l^{\,i}
+b_{8q} (T_8)_i^{\,l}(H_8)_j^{\,i}
       (T_8)_k^{\,j}Q_l^{\,k}.\notag
\end{align}
The redefinition makes the counting of 
independent amplitudes transparent, 
while altering their  physical interpretation. The parameters $a_8$ and $b_8$ are no longer  purely tree-level amplitudes, but also absorb   penguin diagrams and symmetry-breaking effects. The parameters $a_{8q}$ and $b_{8q}$ describe charge-induced symmetry breaking from tree level, whereas $a_{8m}$ and $b_{8m}$ encode mass-induced breaking effects  from both tree- and penguin-level amplitudes.

\begin{table}[t]
\caption{The  amplitudes for   hyperon radiative decays. }
\label{tab:amplitudes}
\begin{ruledtabular}
\begin{tabular}{lc}
Channel & Flavor amplitude\\
\hline
$\Sigma^+\to p\gamma$ &
$\displaystyle \frac{\sin\theta_C}{12}(3b_8+2b_{8q}-b_{8m})$\\[3pt]
$\Lambda\to n\gamma$ &
$\displaystyle \frac{\sin\theta_C}{12\sqrt6}
(-6a_8+3b_8+2a_{8q}-b_{8q}-4a_{8m}-b_{8m})$\\[3pt]
$\Sigma^0\to n\gamma$ &
$\displaystyle \frac{\sin\theta_C}{12\sqrt2}(-3b_8+b_{8q}+b_{8m})$\\[3pt]
$\Xi^0\to\Lambda\gamma$ &
$\displaystyle \frac{\sin\theta_C}{12\sqrt6}
(3a_8-6b_8-a_{8q}+2b_{8q}-a_{8m}-4b_{8m})$\\[3pt]
$\Xi^0\to\Sigma^0\gamma$ &
$\displaystyle \frac{\sin\theta_C}{12\sqrt2}(-3a_8+a_{8q}+a_{8m})$\\[3pt]
$\Xi^-\to\Sigma^-\gamma$ &
$\displaystyle \frac{\sin\theta_C}{12}(3a_8+2a_{8q}-a_{8m})$
\end{tabular}
\end{ruledtabular}
\end{table}

Expanding Eq.~\eqref{eq:total}, we obtain the amplitudes for all decay channels listed in Table~\ref{tab:amplitudes}. We find that some channels receive only a single contraction: the $a$-type amplitude contributes to $\Xi^{0,-}\to \Sigma^{0,-}\gamma$, whereas the $b$-type amplitude contributes to $\Sigma^{+}\to p\gamma$ and $\Sigma^0\to n\gamma$. All remaining channels receive both $a$- and $b$-type contributions.

\noindent{\bf \textit{Global analysis and discussion} }
Using the amplitudes derived and summarized in Table~\ref{tab:amplitudes}, we can perform a global analysis of all weak radiative hyperon decay channels. For the global analysis, we take the six reduced amplitudes in Eq.~\eqref{eq:total} to be real. 
Since the weak interaction violates parity, each weak radiative decay amplitude contains both parity-conserving and parity-violating components, as defined in Eq.~\eqref{ff}:
\begin{align}
{a}_8
&=\frac{G_F e}{M_i+M_f}\,
\bar u_f\, i\sigma_{\mu\nu}q^\nu\epsilon^\mu
\left(f_{a_8}+g_{a_8}\gamma_5\right)u_i,\notag\\
{b}_8
&=\frac{G_F e}{M_i+M_f}\,
\bar u_f\, i\sigma_{\mu\nu}q^\nu\epsilon^\mu
\left(f_{b_8}+g_{b_8}\gamma_5\right)u_i,
\end{align}
Since the photon asymmetry parameter $\alpha_\gamma$, which lies at the heart of the long-standing puzzle, is highly sensitive to the relative magnitudes of the parity-conserving and parity-violating form factors, their independent content must be specified explicitly.
Although the six amplitudes contain 12 form factors, only eight are independent in our global fit.
 For the leading amplitudes $a_8$ and $b_8$, the corresponding parity-conserving and parity-violating form factors are independent since they 
 are induced by the weak Hamiltonian
  in the exact $SU(3)$ flavor symmetry limit. 
  The same remains true when the symmetry-breaking insertion acts on the weak Hamiltonian, where it can alter the underlying $V-A$ structure.
  In contrast, spectator quark insertions leave the weak Hamiltonian unchanged, relating 
  the corresponding parity-conserving and parity-violating form factors and thereby reducing the number of free parameters.

\begin{table}[htbp!]
\caption{Experimental measurements and fitted values for various decay channels, with the branching fractions expressed in units of $(10^{-3})$.}
\label{tab:data}
\begin{ruledtabular}
\begin{tabular}{lcc}
Observable & ${\rm Exp}$ & ${\rm Our\; work}$\\
\hline
$Br({\Sigma^+\to p\gamma})$ & $0.996(28)$\cite{BESIII:2023fhs} & $0.996(28)$ \\
$\alpha_\gamma({\Sigma^+\to p\gamma})$& $-0.69(5)$\cite{PDG} & $-0.685(35)$\\\hline
$Br({\Lambda\to n\gamma})$ & $0.832(66)$\cite{BESIII:2022rgl} & $0.831(66)$ \\
$\alpha_\gamma({\Lambda\to n\gamma})$& $-0.16(11)$\cite{BESIII:2022rgl} & $-0.17(11)$\\\hline
$Br({\Sigma^0\to n\gamma})$& -& $3.7(3.0)\times 10^{-7}$ \\
$\alpha_\gamma({\Sigma^0\to n\gamma})$& -& $-0.685(35)$\\\hline
$Br({\Xi^0\to\Lambda\gamma})$ & $1.24(7)$\cite{PDG} & $1.241(70)$ \\
$\alpha_\gamma({\Xi^0\to\Lambda\gamma})$& $-0.72(5)$\cite{PDG} & $-0.746(36)$\\\hline
$Br({\Xi^0\to\Sigma^0\gamma})$ & $3.69(24)$\cite{BESIII:2025roh} & $3.68(24)$\\
$\alpha_\gamma({\Xi^0\to\Sigma^0\gamma})$& $-0.807(96)$\cite{BESIII:2025roh} & $-0.716(33)$\\\hline
$Br({\Xi^-\to\Sigma^-\gamma})$ & $0.127(23)$\cite{PDG} & $0.127(23)$\\
$\alpha_\gamma({\Xi^-\to\Sigma^-\gamma})$& $1.0(1.3)$\cite{E761:1993unn} & $-0.716(33)$
\end{tabular}
\end{ruledtabular}
\end{table}

Based on the above discussion, the form factors associated with the amplitudes $a_{8q}$, $b_{8q}$, $a_{8m}$, and $b_{8m}$ can be parameterized as
  \begin{align}
  f_{a_{8x}}&=\eta_{a_{8x}}f_{a_8},&
  g_{a_{8x}}&=\eta_{a_{8x}}g_{a_8},\nonumber\\
  f_{b_{8x}}&=\eta_{b_{8x}}f_{b_8},&
  g_{b_{8x}}&=\eta_{b_{8x}}g_{b_8},\qquad x=q,m.
  \end{align}
  This parameterization assumes that each symmetry-breaking amplitudes preserves  the relative   parity structure of its $SU(3)$-symmetric counterpart,
  namely,
$f_{a_{8x}}/g_{a_{8x}}=f_{a_8}/g_{a_8}$ and
  $f_{b_{8x}}/g_{b_{8x}}=f_{b_8}/g_{b_8}$. Under this minimal assumption, the number of free parameters is reduced to eight:
  \begin{equation}
  f_{a_8}, \; f_{b_8},\; g_{a_8},\; g_{b_8},
  \eta_{a_{8q}},\;\eta_{b_{8q}},\;
  \eta_{a_{8m}},\;\eta_{b_{8m}}.
  \label{eq:fit}
  \end{equation}
Under this assumption, the decay asymmetry parameter $\alpha_\gamma$ is
independent of the terms proportional to $a_{8x}$ and $b_{8x}$. This
leads to the relations
\begin{align}
\alpha_\gamma(\Xi^- \to \Sigma^- \gamma)
&=
\alpha_\gamma(\Xi^0 \to \Sigma^0 \gamma), \notag\\
\alpha_\gamma(\Sigma^+ \to p\gamma)
&=
\alpha_\gamma(\Sigma^0 \to n\gamma).
\end{align}
Although these relations appear to be in disagreement with the current
experimental data, the discrepancy is only approximately $1.3\sigma$
owing to the large experimental uncertainties.
A more precise determination of this asymmetry therefore provide a
stringent test of this assumption.

The relevant experimental values are collected in Table~\ref{tab:data}, comprising a total of ten data points.
Fitting these ten observables with eight parameters yields the best-fit values listed in Table~\ref{tab:fit}, with $\chi^2/d.o.f.=0.98$. The  fitted branching fractions and decay asymmetries are presented in Table~\ref{tab:data}.
\begin{table}[htbp!]
\centering
\caption{Fit results for the parameters in Eq.~(\ref{eq:fit}).
}
\label{tab:fit}
\begin{tabular}{cc}
\hline\hline
Form factors
& $\chi^{2}/\mathrm{d.o.f.}=0.98$
\\
\hline

\multirow{2}{*}{Parity conserving (\(f\))}
& \(f_{a_8}=1.47(88)\)
\\
& \(f_{b_8}=4.4(1.8)\)
\\
\hline

\multirow{2}{*}{Parity violating (\(g\))}
& \(g_{a_8}=-0.62(37)\)
\\
& \(g_{b_8}=-1.75(75)\)
\\
\hline

\multirow{2}{*}{Charge insertion (\(\eta_{8q}\))}
& \(\eta_{a_{8q}}=1.7(1.0)\)
\\
& \(\eta_{b_{8q}}=-2.909(38)\)
\\
\hline

\multirow{2}{*}{Mass insertion (\(\eta_{8m}\))}
& \(\eta_{a_{8m}}=5.6(1.6)\)
\\
& \(\eta_{b_{8m}}=-3.138(58)\)
\\
\hline\hline
\end{tabular}
\end{table}
In the present parametrization, $g_{b_8}$ is an effective parity-violating reduced form factor that absorbs symmetry-breaking and electromagnetic-penguin contributions and is therefore allowed to be nonzero. Its fitted value, $g_{b_8}=-1.75(75)$, provides a possible flavor-symmetry realization of the nonzero parity-violating amplitude required to accommodate the measured $\Sigma^+\to p\gamma$ asymmetry. Since this asymmetry is included among the fitted observables, the result constitutes a consistent accommodation rather than an independent prediction of the asymmetry.

The global fit indicates that $SU(3)$ breaking effects are essential for describing weak radiative hyperon decays.
 Since the parameters $\eta$ are defined as the ratios of the form factors associated with $a_{8x}$ ($b_{8x}$) to those of the  reference amplitudes $a_8$ ($b_8$), their
  fitted values quantify the relative size of the ($x=q,m$)
   contributions. 
   Some fitted breaking terms
  can reach approximately five times the corresponding reference contributions. 
  Because $\eta_{a_{8q}}$ and $\eta_{b_{8q}}$ 
arise solely from tree-level contributions, whereas $\eta_{a_{8m}}$ and $\eta_{b_{8m}}$ receive both tree- and penguin-level contributions, the ratios $r_a=\eta_{a_{8q}}/\eta_{a_{8m}}$ and $r_b=\eta_{b_{8q}}/\eta_{b_{8m}}$ may 
provide quantitative measures of the relative tree-level strength.

Regarding the asymmetry parameter puzzle, an $SU(3)$ analysis  including symmetry-breaking  and electromagnetic-penguin effects can accommodate the large negative measured asymmetry in $\Sigma^+\to p\gamma$ without violating Hara’s theorem. 
For $\Xi^-\to\Sigma^-\gamma$, the fit favors a sizable negative asymmetry, $\alpha_\gamma=-0.716(33)$, whereas the current experimental value is $1.0(1.3)$. Although the two central values have opposite signs, their difference amounts to only about $1.3\sigma$ because of the large experimental uncertainty and therefore does not constitute a statistically significant tension. Since this charged mode is also constrained by Hara’s theorem in the exact $SU(3)$ limit, a more precise measurement would provide a particularly sensitive test of the proposed flavor-symmetry realization of the parity-violating amplitude. \\

\noindent{\bf \textit{Summary} }
In this work, we have developed a broken $SU(3)$ flavor analysis of weak radiative decays of spin-1/2  hyperons that includes current-current and electromagnetic-penguin contributions as well as charge- and mass-induced symmetry breaking. 
Six independent reduced amplitudes describe all decay channels, and a minimal relation between the parity-conserving and parity-violating form factors reduces the analysis to eight real parameters. A global fit to ten observables yields $\chi^2/\mathrm{d.o.f.}=0.98$ and 
reproduces the measured branching fractions and photon asymmetries.
The fitted insertion parameters further demonstrate the essential role of $SU(3)$ breaking effects in these decays.

Most importantly, the framework accommodates the large negative measured value of $\alpha_\gamma(\Sigma^+\to p\gamma)$ while remaining consistent with Hara’s theorem. The theorem constrains only the parity-violating current-current amplitude in the exact $SU(3)$ limit, whereas the effective $g_{b_8}$ can receive symmetry-breaking and electromagnetic-penguin contributions. The fitted nonzero $g_{b_8}$ thus provides a possible flavor-symmetry realization of the nonzero parity-violating amplitude. The framework further favors a sizable negative asymmetry in $\Xi^-\to\Sigma^-\gamma$. Although its sign differs from the current experimental central value, 
the measurement is still subject to a large uncertainty and therefore does not constitute a significant discrepancy.
 A more precise measurement of this Hara-constrained charged mode will provide a critical test of this interpretation.\\

\noindent{\bf \textit{Acknowledgments} }
The work of Jin Sun is supported by IBS under project code IBS-R018-D1. 
The work of Ruilin Zhu is supported by NSFC under Grant No. 12322503.
The work of Zhi-Peng Xing is supported by NSFC under Grant Nos. 12375088 and 12335003.

\bibliographystyle{JHEP}
\bibliography{ref}

\end{document}